# Dimensional Crossover Induced Topological Hall Effect in a Magnetic Topological Insulator

Chang Liu[1,*], Yunyi Zang[1,*], Wei Ruan[1,*], Yan Gong[1], Ke He[1,2,†], Xucun Ma[1,2], Qi-Kun Xue[1,2], Yayu Wang[1,2,†]

[1]*State Key Laboratory of Low Dimensional Quantum Physics, Department of Physics, Tsinghua University, Beijing 100084, P. R. China*

[2]*Collaborative Innovation Center of Quantum Matter, Beijing, China*

*\* These authors contributed equally to this work.*

† Emails: kehe@tsinghua.edu.cn; yayuwang@tsinghua.edu.cn

We report transport studies of Mn-doped $Bi_2Te_3$ topological insulator (TI) films with accurately controlled thickness grown by molecular beam epitaxy. We find that films thicker than 5 quintuple-layer (QL) exhibit the usual anomalous Hall effect for magnetic TIs. When the thickness is reduced to 4 QL, however, characteristic features associated with the topological Hall effect (THE) emerge. More surprisingly, the THE vanishes again when the film thickness is further reduced to 3 QL. Theoretical calculations demonstrate that the coupling between the top and bottom surface states at the dimensional crossover regime stabilizes the magnetic skyrmion structure that is responsible for the THE.

Breaking the time reversal symmetry in TIs by means of magnetic doping or magnetic proximity has been a fruitful route for generating exotic topological quantum states [1-5]. An outstanding example is the quantum anomalous Hall effect (QAHE) realized in magnetic TIs [2,6-8], which unambiguously proved the intrinsic mechanism of anomalous Hall effect (AHE) associated with the Berry curvature in momentum space [9,10]. Another intriguing aspect of magnetic TIs is the real space configuration of the local moments, which may also have unique geometrical or topological properties [11-13]. The entanglement of momentum space and real space topology in magnetic TI can lead to novel quantum phenomena, such as the emergence of skyrmions [12-15], that may have potential applications in spintronics [14].

So far ferromagnetic (FM) order has been achieved in a wide variety of transition metal-doped TIs [2,6-8,16-21] and TI-based heterostructures [4,22,23]. Mn-doped $Bi_2Te_3$ compound represents a particularly interesting magnetic TI system. As revealed by the electric gating experiment on Mn-doped $Bi_2Te_{3-y}Se_y$ nanocrystal [17] and STM experiment on Mn-doped $Bi_2Te_3$ crystal [24], the FM order is weakened with increasing surface carrier density, which is consistent with the proposal of surface Dirac fermions mediated Ruderman-Kittel-Kasuya-Yosida (RKKY) mechanism [25,26]. This scenario enables the creation of novel magnetic structure by manipulating the surface states of TIs. In particular, angle-resolved photoemission spectroscopy (ARPES) studies demonstrate that the electronic structure of a TI crossovers from three-dimensional (3D) to two-dimensional (2D) when the film thickness is sufficiently thin so that the top and bottom surface states hybridize [27]. We anticipate that such dimensionality-dependent surface electronic structure may have profound impact on the magnetism of Mn-$Bi_2Te_3$ because of its surface state-mediated RKKY magnetic coupling.

In this work we grow Mn-$Bi_2Te_3$ ultrathin films with varied thickness by using molecular beam epitaxy (MBE). The precise control of film thickness allows us to cover three distinct regimes of TI band structure: the 3D regime where the top and bottom surfaces are decoupled, the 2D limit where the two surfaces strongly hybridize, and the intermediate crossover regime. We find that Mn-$Bi_2Te_3$ films thicker than 5 QL exhibit the usual AHE for magnetic TIs. When the thickness is reduced to 4 QL, however, characteristic features associated with the THE emerge. More surprisingly, the THE vanishes again when the film thickness is further reduced to 3 QL. Theoretical calculations demonstrate that the coupling between the top and bottom surface states at the dimensional crossover regime stabilizes the magnetic skyrmion structure that is responsible for the THE.

The $(Bi_{0.9}Mn_{0.1})_2Te_3$ TI films were grown by MBE on insulating STO (111) substrates following the method reported previously [28]. The thickness of the film was checked by *ex situ* atomic force microscopy, as described in details in supplementary Session A. Transport measurements of the film were carried out by using standard four-probe lock-in method at low frequency with an excitation current of 0.5 μA. Figure 1(a) displays the schematic side view of the magnetic TI film, in which the red arrows represent the Mn magnetic moments. When the film is in the ultrathin limit, the top and bottom topological surface states extending into the bulk will start to hybridize. The electronic structure and spin chirality of the surface Dirac cones are plotted in Fig. 1(b) for both the decoupled and hybridized regimes. Figure 1(c) presents the temperature dependence of Hall traces measured on an 8 QL $(Bi_{0.9}Mn_{0.1})_2Te_3$ film, which is undoubtedly in the 3D regime. At high temperatures, the Hall resistivity $\rho_{yx}$ has a linear field dependence, indicating the sample is in the paramagnetic (PM) state. As the sample is cooled to low temperature, the $\rho_{yx}$ curve evolves into a well-defined hysteretic loop characteristic of the AHE in FM conductors. Both the coercive field ($H_c$) and the zero magnetic field Hall resistivity ($\rho^0_{yx}$) increase with lowering temperature due to the enhancement of FM order.

We then keep the same $(Bi_{0.9}Mn_{0.1})_2Te_3$ chemical composition but reduce the film thickness by 1 QL each time, and map out the evolution of the Hall effect with dimensionality. Figure 2(a) displays the temperature dependent Hall effect curves of the 5 QL sample, which is very similar to that of the 8 QL. The main quantitative difference is the $\rho^0_{yx}$ value increases from ~160 Ω at 8 QL to more than 400 Ω at 5 QL, reflecting the reduction of bulk state conduction in the thinner sample.

When the film thickness is decreased to 4 QL, however, the Hall effect [Fig. 2(b)] exhibits fundamentally different behavior. For temperatures below the Curie temperature $T_c = 18$ K, an extra Hall resistivity feature appears in addition to the usual AHE loop. When the magnetic field is swept up for either polarity, the $\rho_{yx}$ curve deviates from the usual AHE behavior and develops into a broad hump (illustrated by the green patches) superposing on top of the AHE loops. Above a characteristic field scale $H^T$, as indicated by the black arrow, the Hall curve merges back to the usual AHE loop. This extra feature is absent in the process of decreasing magnetic field. The overall profiles of the Hall effect are highly characteristic of the THE due to a real space geometric phase that acts as an emergent electromagnetic field [12,14,29]. Strikingly similar patterns have been observed in bulk and interfacial magnetic skyrmion systems exemplified by MnSi [30,31], FeGe [32] and $SrRuO_3/SrIrO_3$ interface [33].

It is quite surprising that the THE is induced by simply reducing the $(Bi_{0.9}Mn_{0.1})_2Te_3$ film thickness from 5 QL to 4 QL. It is even more surprising that when we further decrease the film thickness to 3 QL, the THE features vanish completely. As shown in Fig. 2(c), the Hall curves of the 3 QL $(Bi_{0.9}Mn_{0.1})_2Te_3$ film basically recover the conventional square-shaped AHE loops without the extra features associated with the THE. The $\rho^0_{yx}$ value now exceeds 1.5 k$\Omega$ at $T = 1.5$ K due to the further suppression of bulk states conduction.

Besides the complex evolution of the Hall effect with film thickness, the longitudinal resistivity $\rho_{xx}$ also displays highly intriguing behaviors. Figure 2(d)-(f) display $\rho_{xx}$ as a function of magnetic field for the three films with different thickness. The butterfly-shaped magneto resistivity (MR) behaviors in the 5 QL and 3 QL samples are the common behavior for FM materials, as has been observed in Cr doped TI before [2,21,34]. In the 4 QL sample, however, a downward hump appears in the magnetic field regime where the THE exists. This observation suggests that the mechanism behind the THE should give rise to a more conductive longitudinal transport.

To further investigate the nature of the THE, we tune the $E_F$ of the 4 QL $(Bi_{0.9}Mn_{0.1})_2Te_3$ sample by applying a gate voltage, and see its effect on the THE. The overall trend revealed by Fig. 3(a) is that $\rho^0_{yx}$ reaches the maximum at $V_g = 0$ and decreases on either side, indicating that $V_g = 0$ corresponds to the charge neutrality point (this can also be seen from the change of slope for the ordinary Hall effect due to normal carriers). The THE is most pronounced at $V_g = -25$ V, and exists in the whole $V_g$ range. The amplitude of the THE decreases at large $V_g$ at either side, similar to the AHE, but shows an apparent electron-hole asymmetry. The amplitude of the topological Hall resistivity $\rho^T_{yx}$ (defined as the difference between the peak of the hump and the AHE loop) is much larger in the negative $V_g$ side than in the positive $V_g$ side. The gate-tuned Hall traces of the 5 QL and 3 QL films are shown in supplementary Fig. S2. They exhibit the typical AHE behavior that has been reported extensively in magnetic TIs [35], and there is no sign of the THE over the entire $V_g$ range.

Figures 3(b)-3(d) summarize the experimental observations about the AHE and THE in the 4 QL $(Bi_{0.9}Mn_{0.1})_2Te_3$. The two effects occur simultaneously at the Curie temperature $T_c$ = 18 K, as shown in Fig. 3(b). Both $\rho^0_{yx}$ (blue) and $\rho^T_{yx}$ (red) increase with lowering temperature, but the former one increases more rapidly whereas the latter one shows a tendency of saturation below $T$ = 6 K. Figure 3(c) shows that the THE disappears at a magnetic field $H^T$ (red) much higher than $H_c$ (blue). We note that due to the coexistence of the two types of Hall effect, $H_c$ can only be defined in an *ad hoc* manner as when the total Hall resistivity crosses zero. Figure 3(d) shows that $\rho^0_{yx}$ (red) and $\rho^T_{yx}$ (blue) have very similar gate voltage dependence. They both reach the maximum in the $V_g$ range between -25 V and 0 V, and exhibit an apparent electron-hole asymmetry. Therefore, the THE is a robust feature of the 4 QL $(Bi_{0.9}Mn_{0.1})_2Te_3$ and exists over a wide range of temperatures, magnetic fields, and gate voltages. We emphasize that without magnetization measurement, the THE and AHE components cannot be quantitatively separated, and the $\rho^T_{yx}$ value plotted here is only an *ad hoc* definition that qualitatively reflects the strength of the THE. Nevertheless, it does not affect the main conclusion regarding the variation of THE with film thickness.

The THE has been observed recently in a heterostructure of magnetic and nonmagnetic TI [15]. Theoretical simulations have shown that it is due to the formation of Néel-type skyrmions induced by the DM interaction [11,36,37]. The Mn-doped $Bi_2Te_3$ films studied here have a different device structure from the heterostructure, and the THE phenomena are even more perplexing. We can create and annihilate the THE by merely varying the film thickness, which suggests that the magnetic structure is closely related to the dimensionality of the TI film. It has been shown that the topological surface states can mediate an effective DM interaction in magnetic TI, and its strength is sensitive to the surface electronic structure [11]. As has been shown by ARPES experiment, reducing the TI film thickness from the 3D to 2D regime opens a hybridization gap at the Dirac point when the top and bottom surface states start to couple. Our results indicate that the change of surface band structure induced by the dimensional crossover have strong influence on the DM interaction and stability of magnetic skyrmions.

In addition to the strong spin-orbit coupling readily available in TI, another prerequisite for a finite DM term is the breaking of inversion symmetry. This naturally occurs for a TI film grown on a substrate due to different inter-surface conditions at the vacuum and substrate sides, causing a potential difference between the two surfaces as revealed in ARPES experiments [27,38]. To demonstrate the effect of inversion symmetry breaking on the THE, we carry out control experiments on two 4 QL Mn-$Bi_2Te_3$ films with and without a 5 nm Te capping layer (supplementary Session C). We find that the THE completely disappears in the sample with Te capping layer, which presumably has a much reduced potential difference between the two surfaces than the uncapped sample. This sharp contrast clearly demonstrates the close relationship between the inversion symmetry breaking and the THE in 4 QL Mn-Bi$_2$Te$_3$, which is consistent with the picture of magnetic skyrmions due to sizable DM interaction.

In order to give a quantitative explanation for the observed THE, we calculate the DM interaction strength characterized by the total off-diagonal spin susceptibility ($\chi_{zx}$ as an example) of the magnetic TI (supplementary Session D) [11]. Figure 4(a) shows the imaginary part of $\chi_{zx}$ as a function of the hybridization gap ($\Delta \approx 2m$) between the top and bottom surface states for a potential difference $2U = 20$ meV (results for other $U$ values are shown and discussed in supplementary Session G). A sharp susceptibility peak appears at a certain hybridization gap, indicating that the DM term is strongest for a specific film thickness and decays rapidly in both sides. This is highly consistent with our experimental observation that the THE only exists in the 4 QL Mn-Bi$_2$Te$_3$ film and justifies the picture of dimensional crossover induced THE. We then calculate the effect of $E_F$ on the DM term for $\Delta = 50$ meV to simulate the gate voltage dependence. As shown in Fig. 4(b), the DM term reaches a maximum at $E_F = 25$ meV and decreases on either side. This evolution trend coincides with our experimental observation of the gate dependence of the THE shown in Fig. 3(d).

Naively, the physical picture concerning the DM term can be attributed to the combined effect of different spin chirality of the two surfaces and the hybridization between them. For each Dirac surface band, the sign of the DM term is determined by the relative position of $E_F$ to the Dirac point [11]. In thicker samples with decoupled surfaces such as the 5 QL Mn-Bi$_2$Te$_3$, electrons at the $E_F$ have opposite spin chirality for the two surfaces. Therefore, the surface state mediated DM term has opposite direction in the two surfaces, and the formation of skyrmions is frustrated [15]. When the film thickness is reduced to 4 QL, the two surface states hybridize and the electrons at the $E_F$ can have the same spin chirality. The DM term of the two surfaces have the same sign and cause a significant enhancement of the total DM interaction hence the stability of skyrmions. Under such situation, the tuning of $E_F$ towards either side will reduce the strength of DM interaction, which explains the gate voltage dependence of THE. When the film thickness is reduced to 3 QL, the TI film enters the 2D regime with a large hybridization gap. The DM term is significantly reduced in this regime and the skyrmion formation is highly unlikely. This simple picture explains why the skyrmion structure and the associated THE are favored only at the 3D to 2D crossover regime with a specific film thickness.

In addition to the calculation of DM interaction, we also simulate the stability of magnetic skyrmion using the same method as that for the TI heterostructure [15]. To check the validity of our model, we first reproduce the previous results as presented in the supplementary Session H. Then we use the 3D tight binding lattice model to calculate the stability of a magnetic skyrmion in Mn-Bi$_2$Te$_3$ with and without inversion symmetry. To simulate the real situation, the $E_F$ is set slightly above the Dirac point. Three types of magnetic skyrmions are included in the simulation. When the inversion symmetry is preserved ($U = 0$, Fig. S9), the system energy keeps increasing with skyrmion size, implying that the FM order is the stable ground state. When the potential difference $U$ is taken into account, the degeneracy between the Néel1 and Néel2 type skyrmions is lift, and an energy minimum appears for the Néel2 type skyrmion at the size of four sites within the Bi$_2$Te$_3$ film plane. This result confirms our numerical calculation shown above and demonstrates the stability of Néel2 type magnetic skyrmions in 4 QL Mn-Bi$_2$Te$_3$ as schematically illustrated in supplementary Fig. S10.

The dimensional crossover induced skyrmion formation picture can also explain the unusual $\rho_{xx}$ behavior for the 4 QL in the THE regime (Fig. 2e). In other films without skyrmion, all the magnetic moments are polarized to the out of plane direction under external magnetic field. In this case, the surface Dirac cone is gapped everywhere in the sample. In the 4 QL sample, the skyrmion formation in some regimes of the sample causes a rotation of magnetic moments from out of plane to the in plane direction, resulting in an effective reduction of magnetic gap [39], hence a more conductive longitudinal transport as shown by the downward hump in the MR in the field regime where the THE exists.

We note that although our experiments and theoretical simulations strongly suggest the existence of magnetic skyrmions in 4 QL Mn-Bi$_2$Te$_3$, without a direct visualization of skyrmion structure we cannot completely rule out other alternative explanations for the observed thickness-dependent transport properties in ultrathin magnetic TI films. A future investigation that may definitive confirm our picture is to observe the stepwise THE due to the quantization of magnetic flux in a single skyrmion [32]. We estimate the length scale of one single skyrmion in our system to be around 30 nm (supplementary Session I), which is much smaller than our device size of around 400 μm. Further miniaturization of device down to the 100 nm range will be needed to observe the quantization of THE due to skyrmions in our 4 QL magnetic TI.

In summary, we demonstrate the creation and annihilation of skyrmion-induced THE in Mn-doped Bi$_2$Te$_3$ by simply varying the film thickness at the dimensional crossover regime.

This unexpected observation illustrates the intimate relationship between the electronic structure and magnetic order in TIs. The simultaneous existence of real space magnetic skyrmion structure and nontrivial momentum space band topology makes this system a unique platform for realizing novel topological quantum phenomena and functional devices. Especially, the sensitivity of skyrmion stability to the dimensionality of magnetic TI enables accurate tuning of skyrmion-related topological transport behavior, which has been explored extensively in recent years for spintronic applications [40-42].

**Acknowledgements**

This work was supported by the National Natural Science Foundation of China and the Ministry of Science and Technology of China.

**Figure Captions:**

FIG. 1. The Hall effect at various temperatures in 8 QL Mn-$Bi_2Te_3$. (a) Schematic side view of the magnetic TI used in the transport measurement. The red arrows indicate the magnetic moments of Mn, and the extension of surface states into the bulk is displayed by the color gradient. (b) The blue and red arrows represent the opposite spin chirality of the top and bottom surface states when they are decoupled. On the right side is the schematic surface band structure

at the ultrathin limit with hybridization. (c) Temperature dependence of the Hall effect in the 8 QL Mn-Bi$_2$Te$_3$ sample at $V_G$ = 0 V. The arrows mark the direction of magnetic field sweep.

FIG. 2. Temperature dependence of the Hall effect and the magneto resistivity in 5 QL, 4 QL and 3 QL Mn-Bi$_2$Te$_3$. (a) In the 5 QL sample, the Hall effect exhibits the usual AHE behavior commonly observed in FM materials. (b) In the 4 QL sample, an extra contribution to the Hall effect (marked by the green patches) appears simultaneous with the AHE. These broad humps are the characteristic behavior of the THE due to magnetic skyrmions. (c) In the 3 QL sample, the Hall effect recovers the usual squared-shaped AHE loop, and the THE features disappear completely. (d)-(f) The longitudinal resistivity $\rho_{xx}$ as a function of magnetic field in the 5 QL, 4 QL and 3 QL samples. The red and blue arrows mark the direction of magnetic field sweep.

FIG. 3. Evolution of the anomalous Hall effect and topological Hall effect in 4 QL Mn-Bi$_2$Te$_3$. (a) Gate voltage dependence of the Hall effect. The THE exists in the whole $V_g$ range, and shows a maximum value near $V_g$. = -25 V, which is close to the charge neutral point when the slope of normal Hall effect changes sign. (b) Evolution of the anomalous Hall resistivity (blue) $\rho^0_{yx}$ and topological Hall resistivity (red) $\rho^T_{yx}$ with temperature. (c) The phase diagram of the 4 QL Mn-Bi$_2$Te$_3$. The broad area between the $H_C$ and $H^T$ lines represents the regime when magnetic skyrmions are stable. (d) Gate voltage dependence of the AHE $\rho^0_{yx}$ (blue) and THE $\rho^T_{yx}$ (red), both showing a peak near the charge neutral point and electron-hole asymmetry.

FIG. 4. Numerical calculation of the DM interaction and simulation of the skyrmion stability. (a) The DM interaction strength as a function of hybridization gap calculated with an inversion symmetry breaking potential $2U$ = 20 meV. The sharp peak indicates that the DM interaction is significant only for a narrow range of hybridization gap corresponding to a specific film thickness. (b) Calculated DM interaction for various chemical potential with the hybridization gap fixed at $\Delta$ = 50 meV. The behavior is consistent with the gate voltage dependence of the THE shown in Fig. 3(d). (c) The energy of three different types of skyrmion configuration relative to the FM state as a function of skyrmion size. The Néel2 type skyrmion has an energy minimum for size of 4 sites, indicating that it is the most stable configuration in the 4 QL Mn-Bi$_2$Te$_3$ film.

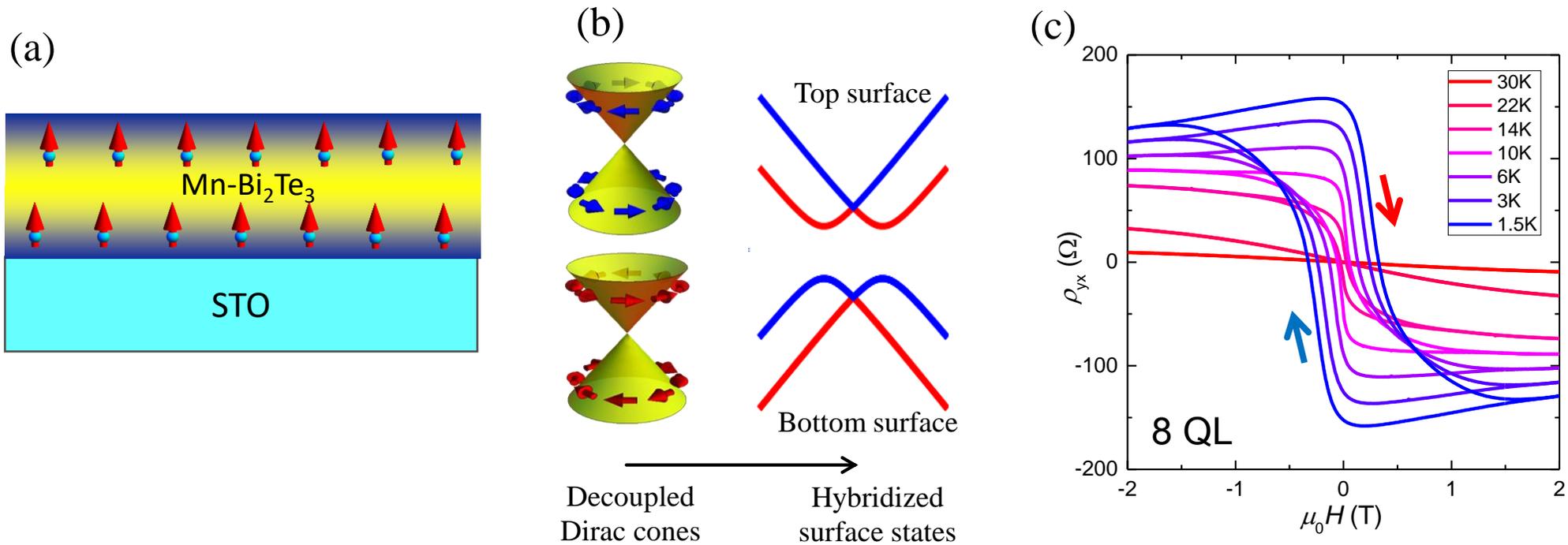

Figure 1

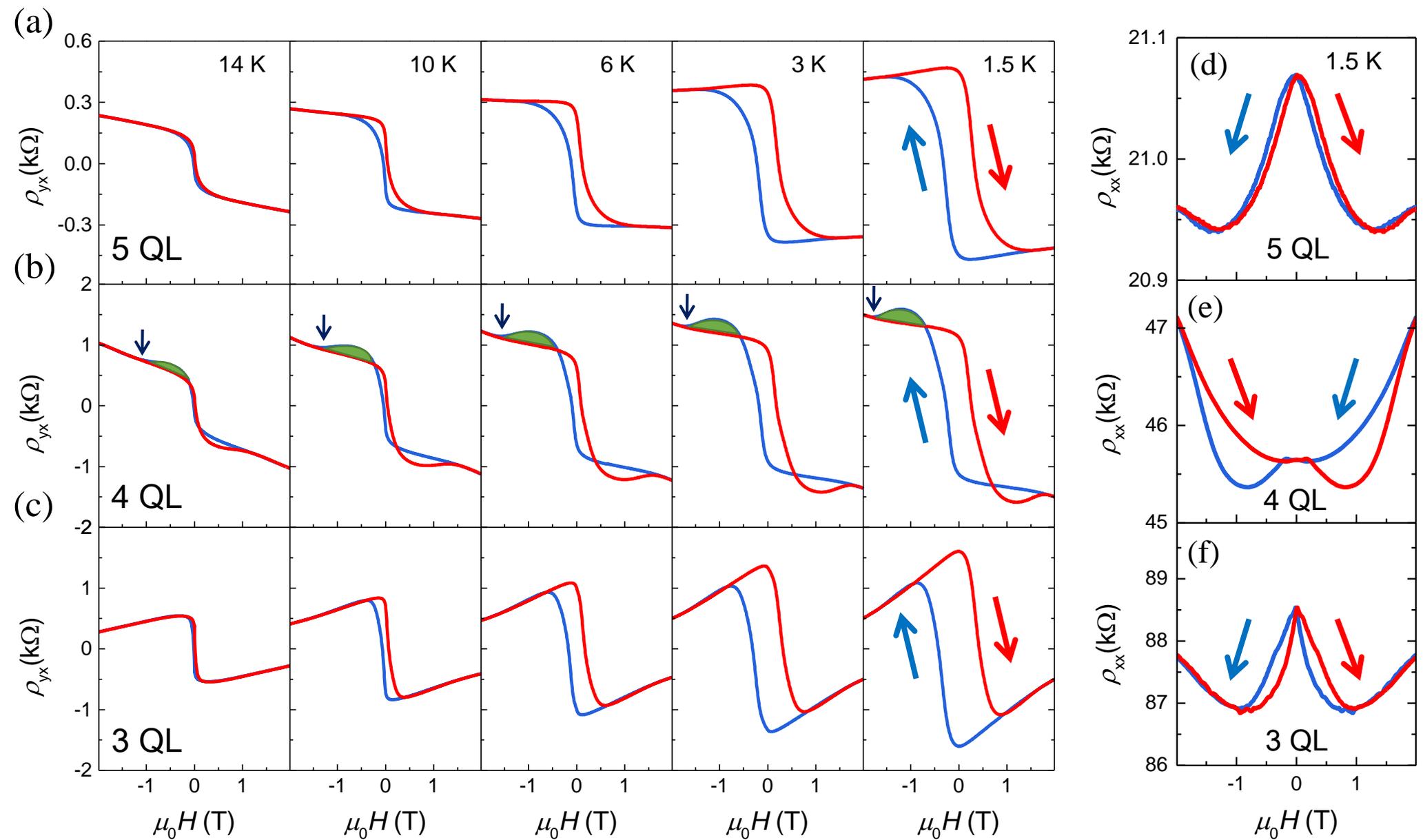

Figure 2

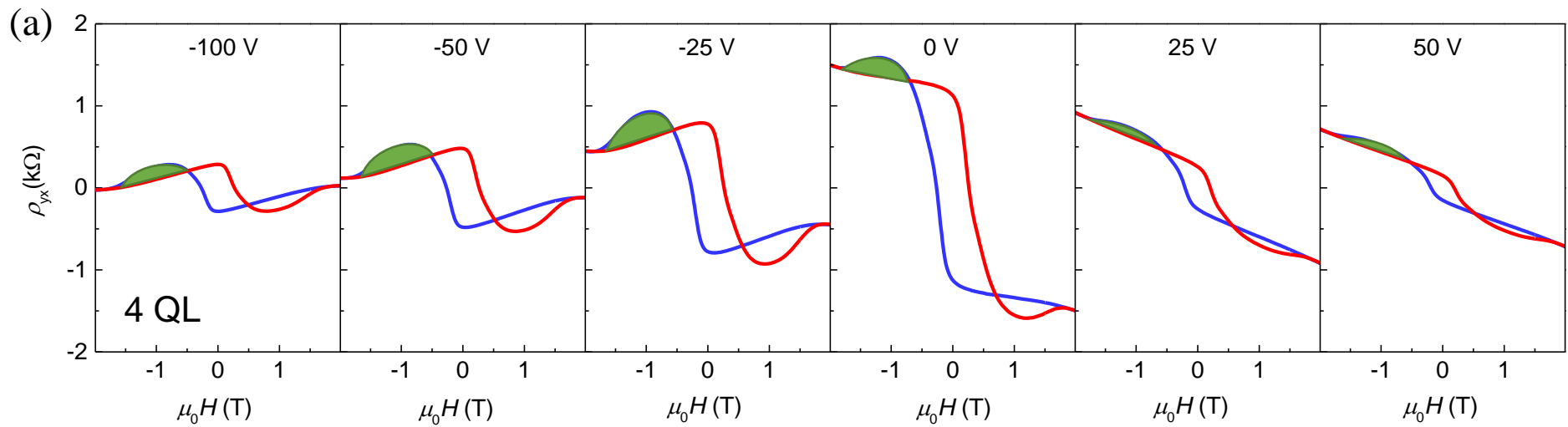
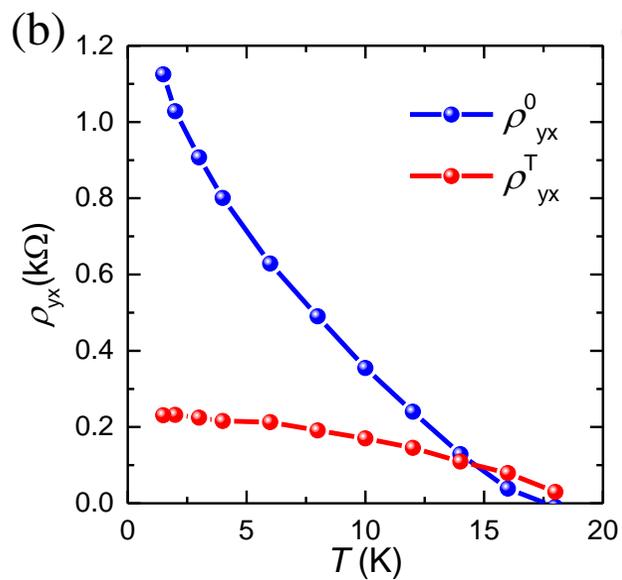
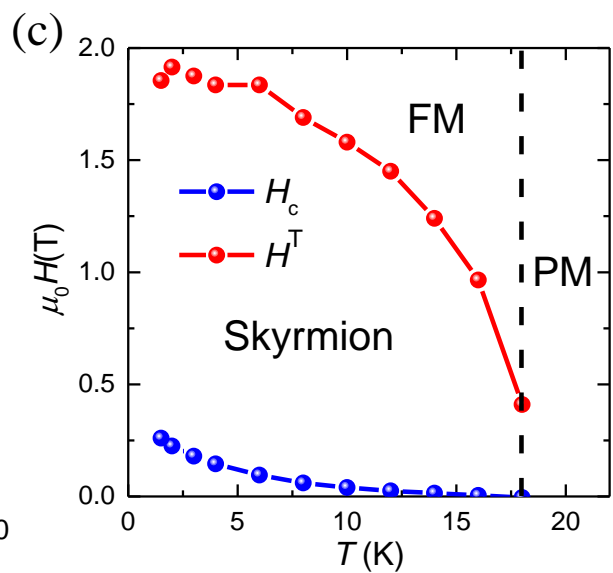
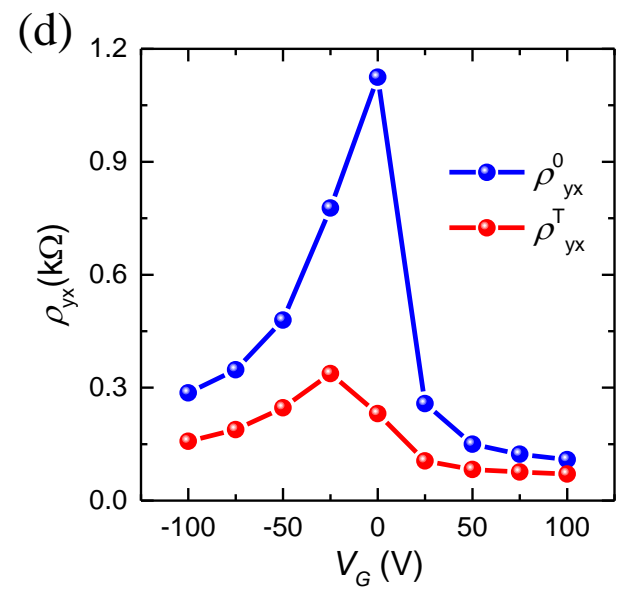

Figure 3

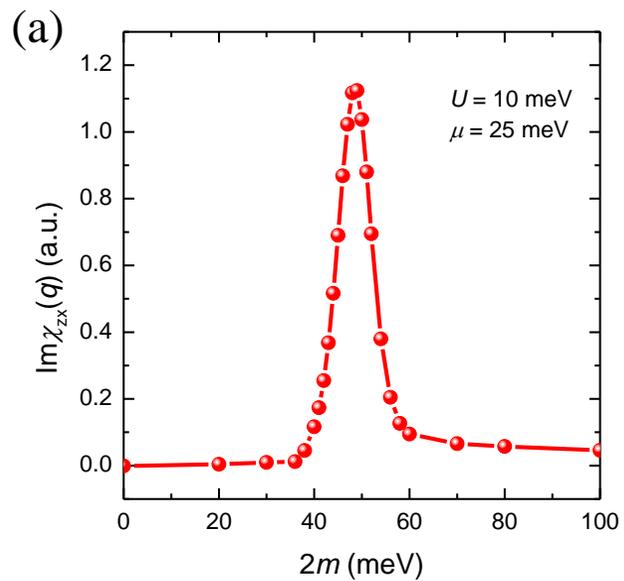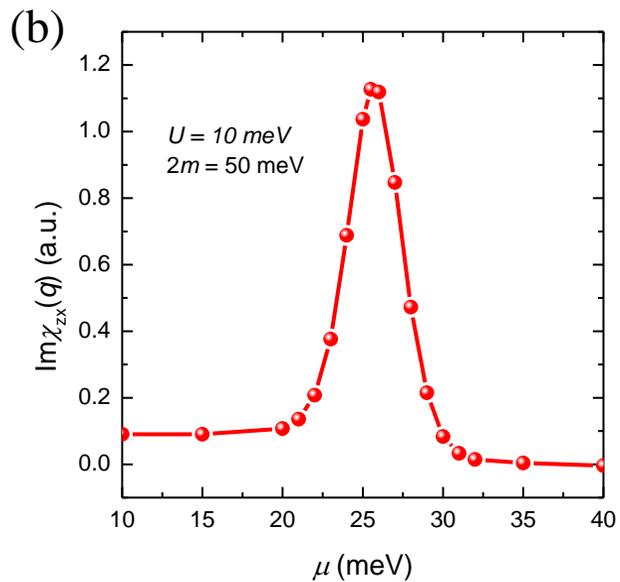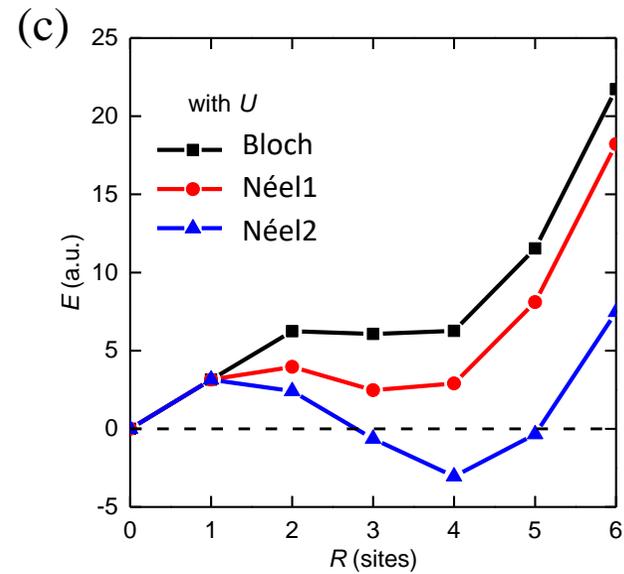

Figure 4

# Supplementary Materials for

# Dimensional Crossover Induced Topological Hall Effect in a Magnetic Topological Insulator

Chang Liu[1,*], Yunyi Zang[1,*], Wei Ruan[1,*], Yan Gong[1], Ke He[1,2 †], Xucun Ma[1,2], Qi-Kun Xue[1,2], Yayu Wang[1,2 †]

[1]*State Key Laboratory of Low Dimensional Quantum Physics, Department of Physics, Tsinghua University, Beijing 100084, P. R. China*

[2]*Collaborative Innovation Center of Quantum Matter, Beijing, China*

*\* These authors contributed equally to this work.*

† Emails: kehe@tsinghua.edu.cn; yayuwang@tsinghua.edu.cn

**Contents:**

SI A: AFM measurements of the film thickness and surface roughness

SI B: Gate voltage dependence of the Hall effect in 5 QL, 4 QL and 3 QL samples

SI C: Tuning Inversion Symmetry by Te capping layer

SI D: Dzyaloshinskii–Moriya interaction for a single Dirac cone

SI E: Numerical results on the spin susceptibility for a single Dirac cone

SI F: Spin susceptibility for hybridized Dirac cones

SI G: Numerical results for hybridized Dirac cones

SI H: Calculation of the magnetic skyrmion stability

SI I: Estimation of the skyrmion density

Figure S1 to S10
References

## SI A: AFM measurements of the film thickness and surface roughness

The thickness of every sample used in our experiment was determined by atomic form microscopy (AFM) after finishing the transport measurement. The overall film quality is checked by reflection high-energy electron diffraction (RHEED) pattern observed during the MBE sample growth. Figure S1a shows the surface morphology of a 4 QL film, which are similar to the typical morphology of magnetic TIs reported previously [1,2]. The nice streaky RHEED pattern shown in the inset reveals the high quality of the sample. The height jump across the sample edge shows that the thickness is around 4 nm (Fig. S1b), indicating that it is indeed a 4 QL film. Figure S1c displays the surface roughness in the middle of the 4 QL sample, in which the root mean squared (rms) roughness is estimated to be 0.33 nm. A line profile along the dashed line is plotted in Fig S1d. It clearly shows that the roughness of the film is around 2 Å.

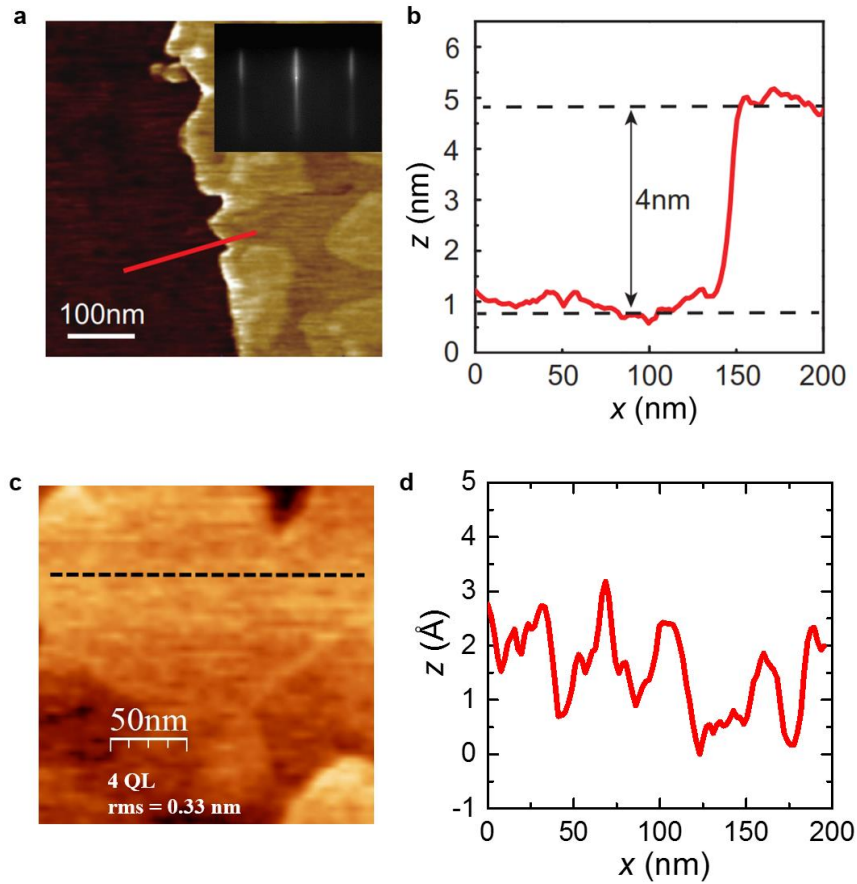

**Figure S1 | AFM determination of film thickness and surface roughness.** (**a**) The AFM image shows the surface morphology near the edge of a 4 QL sample. The inset is the RHEED pattern observed during sample growth. (**b**) The height jump across the sample edge indicates that the

thickness is around 4 nm. (**c**) Surface roughness distribution of the 4 QL film and the surface rms value is around 0.33 nm. (**d**) A line profile taken in the middle of the sample along the direction of the dashed line in figure (**c**).

## SI B: Gate voltage dependence of the Hall effect in 5 QL, 4 QL and 3 QL samples

The gate voltage ($V_g$) dependence of the Hall effect for the Mn-Bi$_2$Te$_3$ samples with three different thicknesses are displayed in Fig. S2. The carrier type of all the three samples can be tuned from electron-like to hole-like by gate, as indiated by the slope of the ordinary Hall effect. For the 3 QL sample, there is no data in the $V_g$ range between 20 V to 90 V due to the highly insulating property in this regime. The most important conclusion regarding the topological Hall effect (THE) is that in the 5 QL and 3 QL films, the THE is absent over the entire $V_g$ range. In the 4 QL sample, on the other hand, the THE has the characteristic $V_g$ dependence as discussed in the main text.

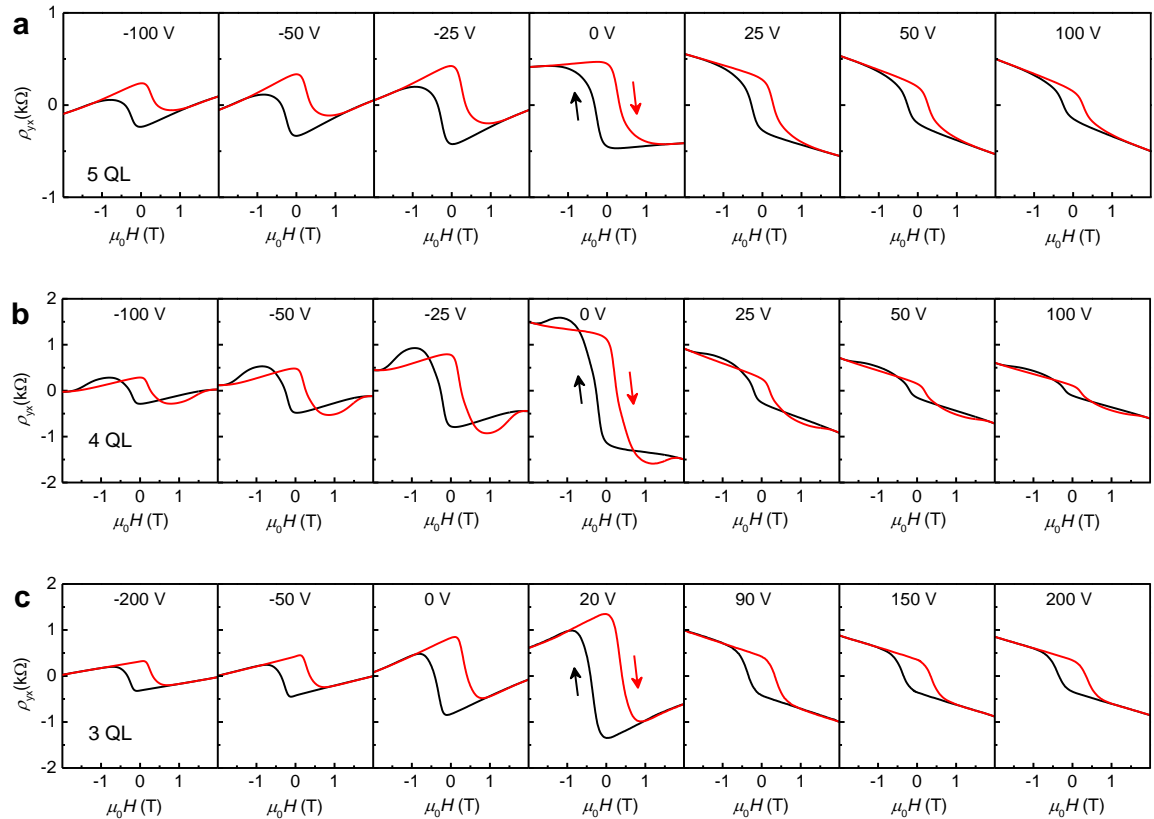

**Figure S2 | Gate voltage dependence of the Hall effect in 5 QL, 4 QL and 3 QL Mn-Bi$_2$Te$_3$**

**samples. (a)**, In the 5 QL the THE is absent over the entire gate voltage range. **(b)**, In the 4 QL sample the THE has the characteristic $V_g$ dependence as discussed in the main text. **(c)**,The THE disappears totally when the thickness is reduced to 3 QL.

### SI C: Tuning Inversion Symmetry by Te capping layer

The temperature dependence of the Hall effect for the 4 QL Mn-Bi$_2$Te$_3$ samples with (upper panel) and without (lower panel) the 5 nm Te capping layer are displayed in Fig. S3. In the sample with Te capping layer on the top surface, the THE totally disappears in the entire temperature regime. Except for the topological Hall term, the Hall effect of the two samples share the same temperature dependent behavior. Figure S4 plots the coercive field (defined in the main text) as a function of temperature for the two samples. It clearly shows that the strength of coercive field is independent of the capping layer, indicating that the capping layer only changes the asymmetry between the top and bottom surfaces without altering the ferromagnetism. The sharp contrast beween the two samples clearly demonstrates the effect on skyrmion formaion induced by inversion symmetry breaking between the two surfaces.

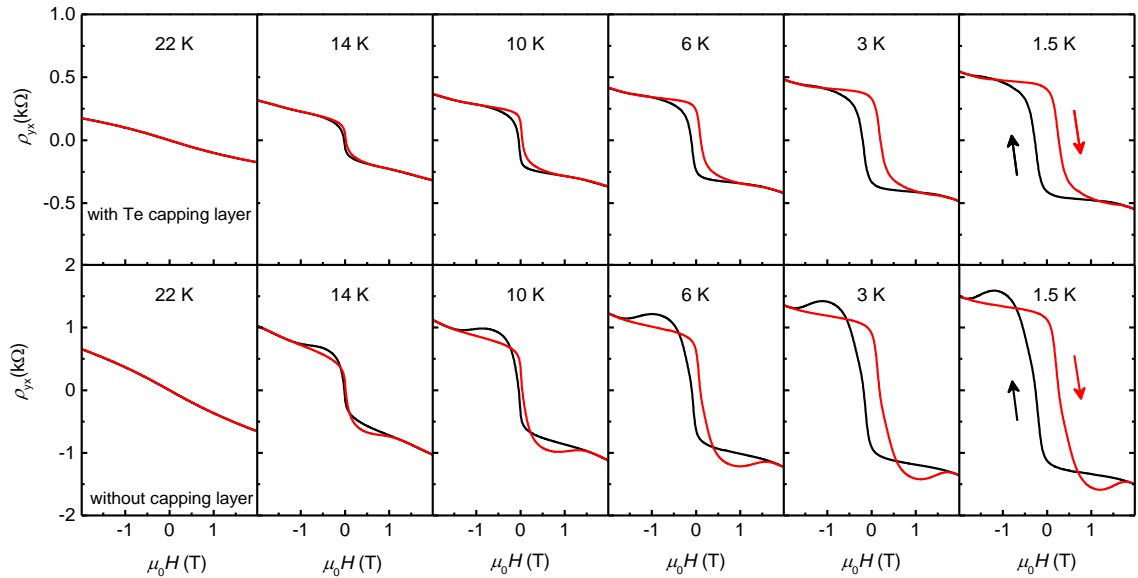

**Figure S3 | Temperature dependence of the Hall effect for the 4 QL Mn-Bi$_2$Te$_3$ samples with (upper panel) and without (lower panel) the 5 nm Te capping layer.** In the sample with Te capping layer on the top surface, the THE disappears and only the usual AHE remains.

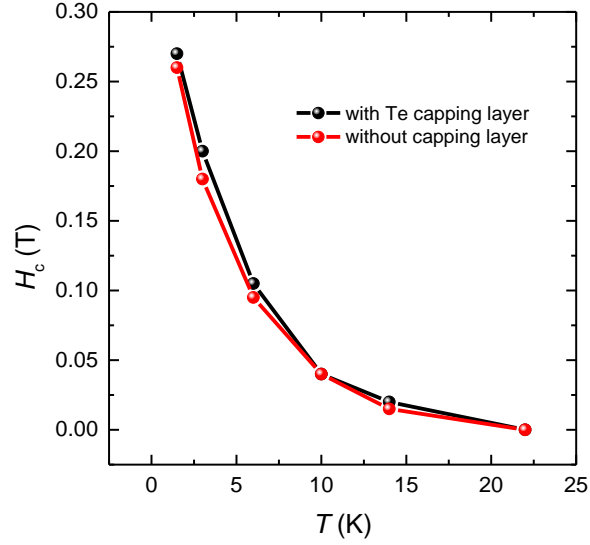

**Figure S4 | Dependence of coercive field with temperature for the 4 QL Mn-Bi$_2$Te$_3$ samples with and without Te capping layer.**

### SI D: Dzyaloshinskii–Moriya interaction for a single Dirac cone

The analytical results presented in this section were obtained in Ref. [5]. We briefly introduce them here because they are essential to our calculations on the hybridized top and bottom surface states in Sec. F, which will be the main focus of this work. In the following Sec. E, we will apply numerical treatment to the single-cone problem as an example, to demonstrate the validity of our numerical methods.

The model Hamiltonian for a 3D topological insulator (TI) in the presence of magnetic exchange coupling can be written as:

$$H_{3D} = v_F \vec{k} \cdot \vec{\sigma} \tau_x + M(k)\tau_z + (J_0 + J_3 \tau_z)\vec{n} \cdot \vec{\sigma}, \quad (1)$$

where $\vec{\sigma}$ and $\vec{\tau}$ are Pauli matrices for spin and orbital degrees of freedom, respectively. $\vec{n}$ is the unit vector representing local magnetic moments of the doped Mn ions. The $J$ terms are magnetic exchange couplings of the surface state electrons with local moments, and $J_3$ term dictates the orbital-dependence of this coupling [3].

When projected onto the 2D surface states, the above Hamiltonian becomes:

$$H_{2D} = \pm v_F(k_x \sigma_y - k_y \sigma_x) + J_3(n_x \sigma_x + n_y \sigma_y) + J_0 n_z \sigma_z, \quad (2)$$

where $+/-$ corresponds to top/bottom surfaces, respectively.

Because the DM (Dzyaloshinskii–Moriya) interaction originates from surface state mediated RKKY interaction, it can be calculated via the spin susceptibility. The effective action when integrating out the fermionic degrees of freedom reads

$$S_{\text{eff}} = \frac{1}{2}\sum_{\alpha\beta} J_\alpha J_\beta \sum_{\vec{q}} n_\alpha(\vec{q})\chi_{\alpha\beta}(\vec{q},0)n_\beta(-\vec{q}). \quad (\alpha,\beta = x,y,z) \tag{3}$$

The DM coupling strength $D$ is determined by the off-diagonal spin susceptibility as $D \sim J_\alpha J_\beta \chi_{\alpha\beta}/2$ with $\alpha \neq \beta$. The spin susceptibility in one-loop approximation is

$$\chi_{\alpha\beta}(\vec{q},i\omega_l) = \frac{T}{V}\sum_{i\omega_n,\vec{k}} \text{tr}[G_0(\vec{k}+\vec{q},i\omega_n+i\omega_l)\sigma_\alpha G_0(\vec{k},i\omega_n)\sigma_\beta], \tag{4}$$

or in the thermodynamical limit

$$\chi_{\alpha\beta}(\vec{q},i\omega_l) = \frac{1}{(2\pi)^2}\int d^2\vec{k}\left[T\sum_{i\omega_n}\text{tr}[G_0(\vec{k}+\vec{q},i\omega_n+i\omega_l)\sigma_\alpha G_0(\vec{k},i\omega_n)\sigma_\beta]\right], \tag{5}$$

where the free Green's function takes the form

$$G_0(\vec{k},i\omega_n) = \frac{1}{i\omega_n - (H_0 - \mu)} = \frac{1}{i\omega_n - v_F(k_x\sigma_y - k_y\sigma_x) - m\sigma_z + \mu}. \tag{6}$$

Here without the loss of generality, a Zeeman term $m\sigma_z$ is added to the free Hamiltonian $H_0$ so that $m$ characterizes the energy gap opened at the Dirac point.

In the $T = 0$ and long wavelength limit, the results for the nonvanishing off-diagonal elements are

$$\chi_{zx}(\vec{q},0) = -\chi_{xz}(\vec{q},0) = -iDq_x, \quad \chi_{zy}(\vec{q},0) = -\chi_{yz}(\vec{q},0) = -iDq_y, \tag{7}$$

where

$$D = \frac{1}{4\pi v_F}[\theta(|m|-\mu) - \theta(|m|+\mu)]. \tag{8}$$

Taking $\chi_{zx}$ as an example, we have

$$\chi_{zx}(\vec{q},0) = \begin{cases} \dfrac{iq_x}{4\pi v_F}, & \mu < -m, \\ 0, & \mu \in (-m,m), \\ -\dfrac{iq_x}{4\pi v_F}, & \mu > m. \end{cases} \tag{9}$$

The above results have a direct implication on the DM interaction strength $D$: $D(\mu) =$

$-D(-\mu)$, and $D(\mu < |m|) = 0$. When the Fermi level ($E_F$) lies in the surface state gap, there will be no DM interaction. The DM interaction is finite only when the $E_F$ cuts through one of the surface bands. It has opposite sign when the $E_F$ cuts through the upper or the lower Dirac cone.

### SI E: Numerical results on the spin susceptibility for a single Dirac cone

We now numerically evaluate the spin susceptibility according to Eq. (5). For simplicity, we take $v_F = m = 1$ in our free Hamiltonian $H_0$ for a single Dirac cone. The temperature $T$ is chosen to be $0.01$ to simulate the zero-$T$ limit, and the summation over the fermionic Matsubara frequencies $\omega_n = (2n + 1)\pi T$ is carried out in the range $-500 \leq n \leq 500$, giving the energy cutoff $\Lambda_\omega = |\omega_{max}| \approx 30$. The integral over the momentum $\vec{k}$ is taken in the range $-5 \leq k_x, k_y \leq 5$, giving a energy cutoff $\Lambda_k \sim v_F|\vec{k}| \approx 5$, much larger than the half gap $m = 1$, which is large enough to capture essential low energy physics in our case. We calculate $\chi_{zx}$ as an example.

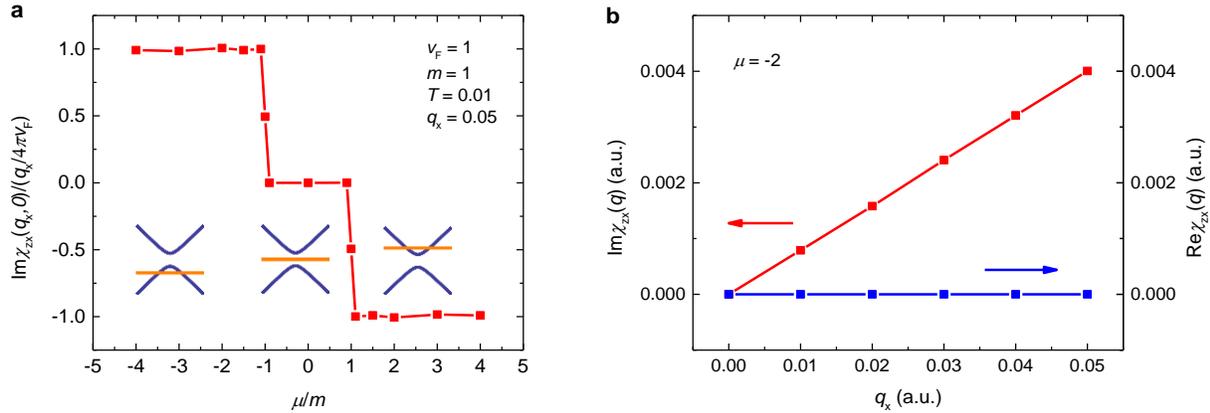

**Figure S5 | Numerical calculation of the spin susceptibility $\chi_{zx}(\vec{q}, 0)$ of a single Dirac cone.** (**a**), Dependence of reduced imaginary spin susceptibility $\mathrm{Im}\chi_{zx}(\vec{q}, 0)/(q_x/4\pi v_F)$ on the reduced chemical potential $\mu/m$. The parameters used are listed in the figure. (**b**), $\chi(q_x, q_y = 0, i\omega_l = 0)$ as a function of $q_x$, which is linear. The chemical potential $\mu$ is set to $-2$ with all the other parameters the same as in (a).

The numerical results are shown in Fig. S5. The real part of $\chi_{zx}$ vanishes as expected from analytical results, which is not shown in Fig. S5(a) but shown in Fig. S5(b) (in blue) as an example. It can be seen from Fig. S5(a) that the imaginary part $\mathrm{Im}\chi_{zx}$ shows two steps, as expected from the fact that $\chi = 0$ with $|\mu| < m$ in the gap, and $\chi$ takes opposite constant values with $\mu$ on opposite sides of the gap. The absolute value of the reduced imaginary part of spin susceptibility reaches unity with the chemical potential $\mu$ cutting through either surface band, which confirms the analytically results quantitatively.

The linear dependence of $\chi_{zx}$ in $q_x$ shown in Fig. S5(b), consistent with Eq. (7), is responsible for the following continuous form of DM interaction Hamiltonian as derived in Ref. [3],

$$H_{DM} = D \int dx dy \left[ \left( n_z \frac{\partial n_x}{\partial x} - n_x \frac{\partial n_z}{\partial x} \right) + \left( n_z \frac{\partial n_y}{\partial y} - n_y \frac{\partial n_z}{\partial y} \right) \right]. \tag{10}$$

**SI F: Spin susceptibility for hybridized Dirac cones**

From Eq. (2), the Hamiltonian for hybridized top and bottom topological surface states in a TI thin film can be modeled as [4]:

$$H_0 = \begin{pmatrix} v_F(k_x \sigma_y - k_y \sigma_x) + U & m \\ m & -v_F(k_x \sigma_y - k_y \sigma_x) - U \end{pmatrix} = v_F(k_x \sigma_y - k_y \sigma_x)\tau_z + m\tau_x + U\tau_z, \tag{11}$$

where $\tau$ is the Pauli matrix for the top/bottom surface state basis $|t\rangle$ and $|b\rangle$, $m$ is the hybridization strength due to the wave function overlap of the top and bottom surface states, and $2U$ is the potential difference between the top and bottom surfaces, which introduces inversion symmetry breaking.

The spin operators for the surface states are

$$\Sigma_\alpha = \begin{pmatrix} \sigma_\alpha & \\ & \sigma_\alpha \end{pmatrix} = \sigma_\alpha^t + \sigma_\alpha^b, \quad \sigma_\alpha^t = \begin{pmatrix} \sigma_\alpha & \\ & \end{pmatrix}, \sigma_\alpha^b = \begin{pmatrix} & \\ & \sigma_\alpha \end{pmatrix}. \tag{12}$$

Where $t/b$ stands for top/bottom. Let the local moments on top/bottom surfaces be $n^{t/b}(\vec{r})$, $J_x = J_y = J_3$ and $J_z = J_0$, the magnetic exchange coupling can be written as

$$H_m = \sum_\alpha J_\alpha (n_\alpha^t \sigma_\alpha^t + n_\alpha^b \sigma_\alpha^b). \tag{13}$$

Integrating out the fermionic degrees of freedom yields the effective action

$$S_{eff} = \frac{1}{2} \sum_{\alpha\beta=x,y,z} \sum_{ij=t,b} J_\alpha J_\beta \sum_{\vec{q}} n_\alpha^i(\vec{q}) \chi_{\alpha\beta}^{ij}(\vec{q}, 0) n_\beta^j(-\vec{q}), \tag{14}$$

where the spin susceptibility is

$$\chi_{\alpha\beta}^{ij}(\vec{q},i\omega_l) = \frac{1}{(2\pi)^2}\int d^2\vec{k}\left[T\sum_{i\omega_n} \text{tr}[G_0(\vec{k}+\vec{q},i\omega_n+i\omega_l)\sigma_\alpha^i G_0(\vec{k},i\omega_n)\sigma_\beta^j]\right]. \quad (15)$$

In order for a magnetic skyrmion configuration to be stablized, the local moment configuration $\vec{n}^t$ and $\vec{n}^b$ should be the same, otherwise it will introduce frustrated local moment configuration [5]. Assuming $\vec{n}^t = \vec{n}^b = \vec{n}$, we have the same effective action as Eq. (3), with the spin susceptibility replaced by

$$\chi_{\alpha\beta}(\vec{q},i\omega_l) = \sum_{ij}\chi_{\alpha\beta}^{ij}(\vec{q},i\omega_l) = \frac{1}{(2\pi)^2}\int d^2\vec{k}\left[T\sum_{i\omega_n} \text{tr}[G_0(\vec{k}+\vec{q},i\omega_n+i\omega_l)\Sigma_\alpha G_0(\vec{k},i\omega_n)\Sigma_\beta]\right], \quad (16)$$

and the free Green's function

$$G_0(\vec{k},i\omega_n) = \frac{1}{i\omega_n - (H_0-\mu)} = \frac{1}{i\omega_n - v_F(k_x\sigma_y - k_y\sigma_x)\tau_z - m\tau_x - U\tau_z + \mu} \quad (17)$$

### SI G: Numerical results for hybridized Dirac cones

We then perform numerical calculations on the spin susceptibility $\chi_{zx}(\vec{q},0)$ for hybridized top and bottom surface Dirac cones, based on Eq. (16). We take $v_F = 3.32\times 10^5$ m/s [6] and $\mu = 25$ meV. From ARPES measurements and first-principles calculations for $Bi_2Te_3$, the Dirac point of the topological surface states sinks below the top of the bulk valance band by tens of meV [6]. As in our experiments, the Hall resistance displays both *n*- and *p*-type behavior (see Fig. S2) around the charge neutral point, therefore the $E_F$ should be almost pinned at the top of bulk valance band and we set $\mu$ to be 25 meV above the Dirac point, to be in the same order of magnitude as the experimental value.

The temperature is chosen to be $T = 1$ meV to approximate the zero-temperature limit, and we sum over the fermionic Matsubara frequencies $\omega_n = (2n+1)\pi T$ in the range $-500 \le n \le$

500, giving the energy cutoff $\Lambda_\omega = |\omega_{max}| \approx 3000$ meV. The integral over momentum $\vec{k}$ is taken in the range $|k_x|, |k_y| \leq 0.1$ Å$^{-1}$, which roughly approximates the momentum space boundary of the surface states [6].

For a fixed wave vector $q_x = 0.001$ Å$^{-1}$, $q_y = 0$, the imaginary part of the off-diagonal spin susceptibility $\chi_{zx}$ for different potential differences $U$ and hybridization strengths $m$ is summarized in Fig. S6, while the real part vanishes for all parameters so is not shown in the figure. One striking feature revealed by this result is that $\text{Im}\chi_{zx}$ exhibits a pronouced peak centered around a specific hybridization strength $m$ close to the chemical potential $\mu = 25$ meV, for nonvanishing $U$ values. When $U = 0$, there is no inversion symmetry breaking, so $\text{Im}\chi_{zx} \equiv 0$, i.e. no DM interaction is present [7]. While with finite $U$, DM interaction is possible for intermediate values of $m$, which is tunned by the thickness of the TI thin film. With larger or smaller $m$ corresponding to thinner or thicker films, $\text{Im}\chi_{zx}$ or the DM interaction vanishes.

The $\vec{q}$-dependence of $\chi_{zx}$ is also examined and summarized in Fig. S7, which demonstrate similar $\vec{q}$ dependence as in Eq. (7):

$$\chi_{zx}(\vec{q}, 0) \propto iq_x. \tag{18}$$

The results for $\chi_{xz}$, $\chi_{zy}$ and $\chi_{yz}$ can be obtained by symmetry considerations.

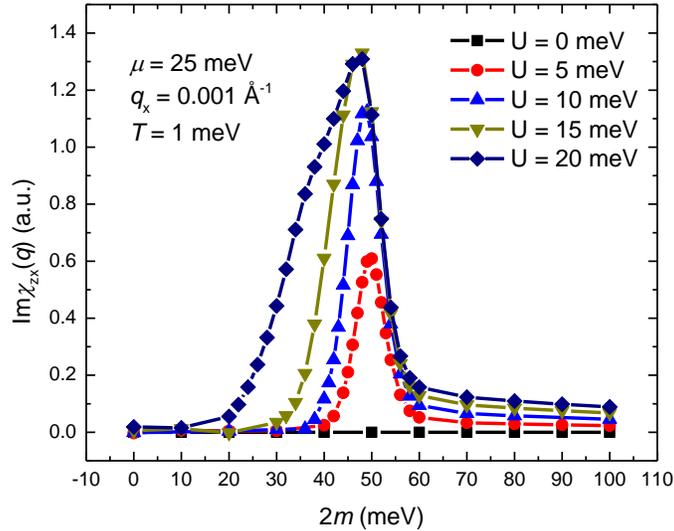

**Figure S6 | Imaginary part of the spin susceptibility $\text{Im}\chi_{zx}(\vec{q}, 0)$ as a function of the hybridization strength $m$ for different values of potential difference $U$.**

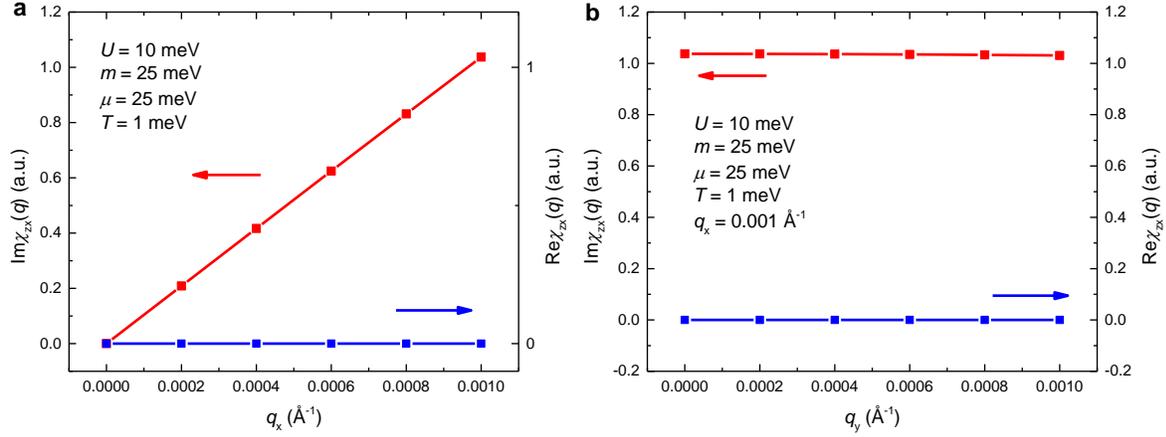

**Figure S7 | Spin susceptibility $\chi_{zx}(\vec{q}, 0)$ as a function of wavevector $\vec{q}$.** Both the real part (blue curves, right axes) and the imaginary part (red curves, left axes) are shown. **(a)** $q_x$-dependence of $\chi$ with $q_y = 0$. **(b)** $q_y$-dependence of $\chi$ with $q_x$ fixed at 0.001 Å$^{-1}$.

From Eq. (14) and Eq. (15), we can see that the total DM interaction actually contains four different contributions: (1) exchange ($\chi^{tt}$) between two local moments on the top surface; (2) exchange ($\chi^{bb}$) between two local moments on the bottom surface; (3)(4) exchange ($\chi^{tb}$) between one local moment on the top surface and one on the bottom surface via the hybridization of both surface states, and vice versa ($\chi^{bt}$). To identify these contributions, we choose $U = 10$ meV as an example, which is close to our assumed physical values. The peak structure in Fig. S6 at intermediate values of $m$ can be ascribed to two reasons, as shown in Fig. S8:

1. Sum of the exchange interactions between top-top local moments $\chi^{tt}$ and bottom-bottom local moments $\chi^{bb}$. In the small $m$ limit, the $E_F$ cuts through the cones with the opposite chiralities, so that the two exchange terms cancel each other. While in the large m limit, the $E_F$ lies in the opened gap of both the top and bottom surface states, so that both exchange terms are vanishingly small.

2. The exchange between top surface local moment and a bottom one $\chi^{tb}(\chi^{bt})$. Both terms show a peak when the $E_F$ lies at the gap edge of the gapped Dirac cones, because the wavefunction overlap between the top surface state and the bottom surface state takes maximum value at the gap edge.

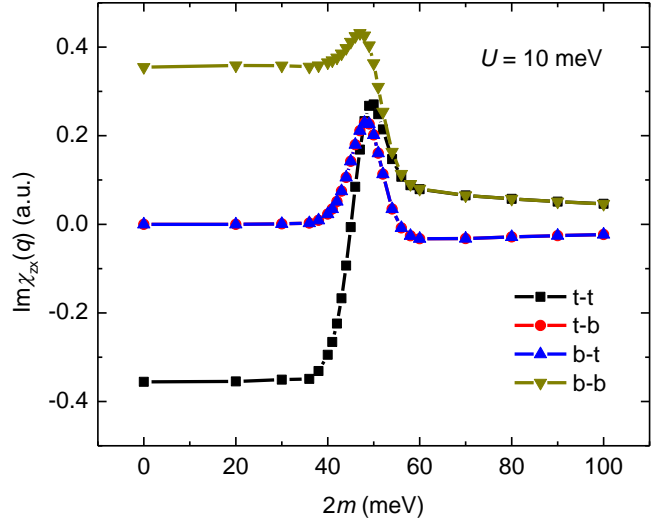

**Figure S8 | Spin susceptibility $\chi_{zx}^{ij}(\vec{q},0)$ as a function of the wavevector $\vec{q}$ for $i,j = t, b$.** For example, $\chi^{tt}$ mediates the exchange between local moments on the top surface, and $\chi^{tb}$ mediates the exchange between one local moment from the top and one from the bottom surface.

### SI H: Calculation of the magnetic skyrmion stability

From the results of $\chi$ calculations, we obtained the dependence of the DM interaction strength on the film thickness and chemical potential, which is consistent with our experimental findings. Now we carry out lattice model calculations to demonstrate the stability of magnetic skyrmion configurations in the presence of a surface potential difference $U$ in a magnetic TI thin film. Our starting point is the lattice model corresponding to the tight-binding Hamiltonian for a 3D TI [5]:

$$H_0 = t \sum_{\alpha=x,y,z} \sin k_\alpha \sigma_\alpha \tau_x + [m + 2D_1 \sum_{\alpha=x,y,z} (1 - \cos k_\alpha)]\tau_z, \tag{19}$$

plus an exchange between electrons and the doped local moments, as well as a surface potential difference term:

$$H' = -(J_0 - J_3 \tau_z)\vec{n} \cdot \vec{\sigma} + \left(-U + \frac{2zU}{N_z - 1}\right), \tag{20}$$

where $\sigma$ and $\tau$ are pauli matrices for the spin and orbital degrees of freedom respectively. $J_3$ accounts for the asymmetric part of exchange coupling [5]. The linear potential term is added so that it takes values of $\pm U$ on the top and bottom surfaces respectively.

In our calculation, we take $N_x = N_y = 16$ sites and $N_z = 4$ layers, and impose the periodic boundary condition in both x and y directions. The band structure parameters are set as $t = 1$, $m = -1$ and $D_1 = 1$ to realize the strong TI phase. The values for exchange interactions are $J_0 = 0.1$ and $J_3 = 0.02$. The potential difference is $U = 0.05$. We set our chemical potential slighly above the Dirac point to simulate the real materials under investigations. Due to the finite size effect, these values are chosen not to correspond to real physical values, but to give qualitative demonstration of the skyrmion stability. The band structure parameters and the exchange coupling strengths are indeed the same as in Ref. [5], which were proved to give qualitative understanding of the skyrmion stability.

As in Ref. [5], we have also considered three types of skyrmion conifgurations, namely the Bloch, Néel1, and Néel2 types. They are expressed in a general form:

$$\begin{cases} n_x = \sin\left[\pi\left(1 - \frac{r}{R}\right)\right]\cos(\theta + \phi), \\ n_y = \sin\left[\pi\left(1 - \frac{r}{R}\right)\right]\sin(\theta + \phi), \\ n_z = \cos\left[\pi\left(1 - \frac{r}{R}\right)\right], \end{cases} \quad (21)$$

for $r < R$. For $r \geq R$, we take $n_x = n_y = 0$ and $n_z = 1$. $\phi = \pi/2, 0, \pi$ correspond to the Bloch-, Néel1-, and Néel2-type, respectively. The skyrmion radius in the unit of lattice constant is varied from 0 to 6, where 0 corresponds to the ferromagnetic phase. The total energy as a function of $R$ compared with the reference value at $R = 0$ is then calculated.

Figure S9 shows the calculated total energy of skyrmions with different sizes, with the inversion symmetry preserved. The total energy with finite skyrmion size $(R > 0)$ is larger than that in the ferromagnetic case $(R = 0)$, and increases with increasing skyrmion size. This result clearly demonstrates that the magnetic skyrmions are not stable in this case, and is consistent with the results calculated in Ref. [5]. The results with finite $U$ demonstrating the stability of skyrmions are presented in Fig. 4(c) in the main text.

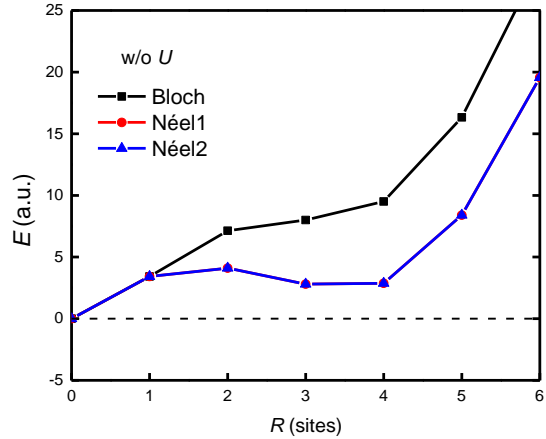

**Figure S9 | Skyrmion stability calculation with $U = 0$.** The calculation shows that no magnetic skyrmion can be stablized when the inversion symmetry is preserved.

### SI I: Estimation of the skyrmion density

In the magnetic skyrmion systems, the magnidude of THE is closely associated with the emergent electricmagnetic field which is determined by the length scale of skyrmions. Therefore, by calculating the effective magnetic field from the magnitude of topological Hall resistivity, we can roughly estimate the skyrmion density. The schematic picture of the Néel2 type skyrmion in our sample is show in Fig S10.

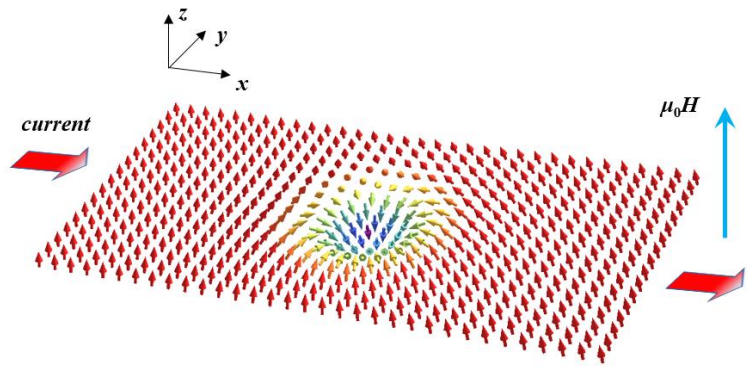

**Figure S10 | Néel2 type skyrmion congifuration.** The Néel2 type skyrmion is distributed in the film $x$-$y$ plane. The current (red arrow) flows in the $x$ direction while the magnetic field is in the $z$ direction.

In magnetic skyrmion systems, each skyrmion traps a magnetic flux quantum $\Phi_0$. The topological Hall resistivity induced by skyrioms can be expressed by [8]

$$\rho^T = PR_0 n_{sk} \phi_0 \qquad (22)$$

where $P$ is the spin polarization of electrons in the films, $R_0$ is the ordinary Hall coefficient, and $n_{sk}$ is skyrmion density. Based on frist principle calculation and point contact Andreev reflection measurement [9,10], V and Cr doped $Bi_2Te_3$ have spin polarization of around 50%. Roughly we can set the spin polarization of Mn doped $Bi_2Te_3$ to be in the same order of magnitude. The ordinary Hall coefficient can be calculated from the slope of the linear part of Hall reisitivity. Using the extracted topological Hall resistivity, the skymion density is estimated to be around $9 \times 10^{14}/m^2$. The estimated length scale of a single skyrmion is $n_{sk}^{-1/2} \sim 30$ nm, which sets the upper limit of the skymion size in our materials. Such value is quite reasonable compared with the typical diameter of DM interaction induced skyrmions, which usually ranges from 5 nm to 100 nm [11,12].